# Modeling Events and Events of Events in Software Engineering

Sabah Al-Fedaghi
Computer Engineering Department
Kuwait University
Kuwait
sabah.alfedaghi@ku.edu.kw

*Abstract*—A model is a simplified representation of portion of reality that hides a system's nonessential characteristics. It provides a means for reducing complexity as well as visualization and communication and a basis for building it. Most models involve graphic languages during many of the software lifecycle phases. A new model, called thinging machine (TM), has recently been developed as an extension of the input-process-output framework. The paper focuses on events in a TM, offering a new perspective that captures a system's dynamic behaviors and a means of diagrammatically modeling events. The event notion is an important factor in giving semantics to specifications and providing a natural way to specify the interfaces and observable behavior of system components. Specifically, five generic TM event processes are analyzed: create, process, receive, release, and transfer. All events can be mapped (or reduced) to the events of these five event processes.

I. INTRODUCTION

The impact of failures in software system development are very costly. Communication and integration difficulties can arise from variations in representations of notions, which have resulted in a multiplicity of schematics, depictions, and representations, creating difficulties in managing processes, as well as inconsistent usage. A key problem in this context is the lack of a unifying theoretical framework for software system development. To avoid potential failures, new models have been proposed to specify a system's structure and behavior. A model is a simplified representation of a portion of reality that hides a system's nonessential characteristics. It provides a means for visualization and communication and a basis for building the system. Most models involve graphic languages during many phases of the software lifecycle [1].

According to Kazman [2], software engineers deal with more complex problems than engineers in any other engineering discipline. Decades of work on software abstraction and specification have resulted in gaining intellectual control over systems of ever-increasing complexity, which has motivated the adoption of a modeling approach throughout the software-development process. Software engineers develop conceptual models using tools such as *the* Universal Modeling Language (UML) and System Modeling Language (SysML). A new model, called the thinging machine (TM), has recently been developed as an extension of the input-process-output framework [3-12].

This paper focuses on events in modeling and offers a new perspective that captures a system's dynamic behaviors. According to Talcot [13], the notion of an event can be interpreted as follows:

- Events can be used to give semantics to specifications and to support runtime observation/monitoring adaptation, security decisions, and trust building. Events support new programming abstractions that deal with actions and interactions rather than state transformations.

- Events are a natural way to think about reactive systems and to specify the interfaces and observable behavior components of open systems. Events can also form the basis for specifying coordination and composition of components [13].

Barga et al. [14] stated that the notion of events must be further explored for real-world tasks such as machine translation, language generation, and document classification. Studying events has produced various technologies, including data-stream management, complex event processing, and asynchronous messaging. Barga et al. [14] observed that "business events are the real drivers of the enterprise today because they represent changes in the state of the business. *Unfortunately*, as in the case of data management in pre-database days, every usage area of business events today tends to build its own special purpose infrastructure to filter, process, and propagate events" (Italic added). Grez et al. [15] declared that complex event processing "has emerged as the unifying field for technologies that require processing and correlating distributed data sources in real-time. . . . However, existing languages lack clear semantics, making them hard to understand and generalize". Grez et al. [15] cited event-processing applications in diverse domains such as network intrusion detection [16], industrial control systems [17], and real-time analytics [18]. According to Singh et al. [19], "The increasing popularity of event-based systems has opened up new challenging issues for them. One such issue is to carry out requirements analysis of event-based systems and build conceptual models."

This paper involves a study of semantic events, as a bridge toward understanding the dynamic behavior of systems. Specifically, the five generic TM event processes are analyzed: create, process, receive, release, and transfer. These processes are basic operations in any system, physical or otherwise. *All events can be mapped (or reduced) to the events of these five processes*.

## II. ABOUT EVENTS

Talcot [13] observed that many notions of an event are utilized for various purposes, including an event as sending or receiving a message, as an action shared by two processes, and as something that is nested or happens over time.

### A. A Glimpse from Philosophy

In questioning the nature of events, the *Stanford Encyclopedia of Philosophy* [20] provides controversial differences between events and physical objects:

- A difference in the mode of being: material objects are said to *exist*, while events are said to *occur*, *happen*, or *take place*.
- Objects are located in space and have vague temporal boundaries Events have temporal boundaries and vague spatial boundaries.
- Objects can move; events cannot.
- Objects exist in time and persist through time; events take up time and persist by having different stages at different times.

Objects are viewed as prime actors in events; objectless events are uncommon. However, eventless objects are also viewed as prime actors in events; events make up the lives of objects [20].

Some philosophers claim that no significant distinction exists between objects and events: an object would simply be a "monotonous" event; an event would be an "unstable" object. Some philosophers also reject the distinction on the side of events, conceiving all entities as processes [20].

Whitehead [21] affirmed that everything is an event. For Whitehead, Cleopatra's Needle on the Victoria Embankment in London is eventful at every moment: "A physicist who looks on that part of the life of nature as a dance of electrons will tell you that daily it has lost some molecules and gained others, and even the plain man can see that it gets dirtier and is occasionally washed" [21]. Thus, components of reality are actual entities/occasions, such as Cleopatra's Needle.

Whitehead [21] used the term "event" to mean a nexus of actual occasions; here, the actual occasion is the limiting type of an event with only one member [22]. An event may be just one particular occasion or a multiplicity of becomings [22]. According to Whitehead [21], "What we discern is the specific character of a place through a period of time. This is what I mean by an 'event.'"

### B. A Glimpse from Computer Science

In computing, software recognizes an event as an occurrence, often originating asynchronously from the external environment. Event sources include a user and a hardware device (e.g., a timer), but a system may also trigger its own events (e.g., communicating the completion of a task) [23].

In software engineering, a software component uses events as objects/messages when it wants to notify other components about a state change. An event model is a software architecture (a set of classes and interfaces) that determines how components can create, trigger, and distribute events [24]. Events can be flows (e.g., data flows) and can occur as a result of timing or at some unpredictable point in time [24]. Examples of events include a customer placing a reservation (flow), an accounting system receiving transaction details (flow), management requesting a weekly report (temporal), and a credit card being verified (control) [25]. According to the site Event Modeling [26], event modeling involves seven steps:

- Documenting all of the events that one can conceive of having happened.
- Creating a plausible story comprising these events so that they are linearly arranged in order.
- Constructing wireframes so that the system's blueprint has the information's sources and destinations.
- Dividing the wireframes into separate swim lanes.
- Identifying inputs.
- Identifying outputs.
- Organizing the events into swim lanes.
- Elaborating upon scenarios.

Accordingly, the event model should account for every "data field" and have an origin and a destination for all information.

Real-time systems are often event-driven (e.g., a landline phone-switching system responds to events such as "receiver off hook" by generating a dial tone). Event-driven models are based on the assumptions that a system has a finite number of states and that events (stimuli) may cause a transition from one state to another. Event-driven models can be created using UML state diagrams [27]. In distributed systems, an event is an asynchronous message that signals the occurrence of situations of interest [28]. A primitive event corresponds to an elementary occurrence, whereas a composite event is a combination of other primitive or composite events (e.g., a sequence, disjunction, conjunction, and negation of events). Events can be either produced or consumed. [28].

In object-oriented methodology, an event is an external stimulus from one object to another that occurs at a particular point in time [1]. It is a transmission of information from one object to another. A scenario is a sequence of events that occurs during one particular execution of a system [1]. According to Singh et al. [19], no object-oriented analysis techniques or tools focus on event-based requirements analysis; rather, they all use behavior-based approaches.

*(IJCSIS) International Journal of Computer Science and Information Security,*
*Vol., 18, No. 1, 2020*The next section discusses the TM model, with emphasis on the thing/machine concept (thimac) used to provide an ontology based on a single unifying element, which is motivated by a desire to integrate object- versus process-based models. Section IV illustrates the TM by applying it in the context of (1) relating events to Davidson's [29] actions and (2) relating events to facts as described in the *Stanford Encyclopedia of Philosophy* [20]. Section V offers a case study of events using the object-oriented methodology [1] and produces an alternative TM solution.

### III. THINGING MACHINE

The TM relies more on Heidegger's [30] notion of *things* than it does on the notion of objects. Heidegger's works on thinging have been applied in various scientific fields (e.g., design thinking [31] and information services [32]). "Almost anything can be labeled with the word thing" [33]. According to Heidegger [30], a thing is self-sustained, self-supporting, or independent—something that stands on its own. Things have unique "thingy Qualities" [34] that are related to reality and therefore not typically found in industrially generated objects. A thing "things" (Heidegger's [30] term); that is, it gathers, unites, or ties together its constituents, in the same way that a bridge unifies aspects of its environment (e.g., a stream, its banks, and the surrounding landscape) [35]. Such a notion is more suitable to our aim than reductive abstraction of entities, as reflected by the term "object".

Building on such an approach, things are combined with the concept of a process by viewing them as blocs called single ontological things/machines, or thimacs, which populate a world that is itself a thimac (we call it a system). Every part of this world is a thimac, forming a thimac-ing network. A unit of such a universe has dual being as a thing and as a machine. A thing is created, processed, released, transferred, and/or received. A machine creates, processes, releases, transfers, and/or receives things. We will alternate between the terms "thimac", "thing", and "machine" according to the context.

The term "thimac" designates what simultaneously divides and brings together a thing and a machine. Every thimac appears in a system either by creation or by importation from outside the system. They are the concomitants (required components) of a system and form the current fixity of being for any system that continuously changes from one form (thing/machine) to another.

A system is the overall constellation thimac that structures all thimacs in the problem under consideration. It provides the problem's unifying element through space and time as integral subthimacs, not as the sum of individual subthimacs. Thimacs inside a system are understood not as things with properties but as ensembles of things and machines that constantly interact with each another and with the out-of-system world.

In this complex model, events appears propagate, and constantly recur in various parts of the system with repeatable occurrences and stable regularities. The whole system in its dynamic state is an event or, more accurately, an event of events. The TM model is a representation (mimesis) of a portion of reality. It is an expression of information, solutions, knowledge concerning the system, system architecture, system functionality, business processes, etc.

Accordingly, a thimac's existence depends on its position in the larger system, as either a thing that flows in other machines or a machine that handles a flow of things (create, process, release, transfer, and receive). It brings together and embraces both "thingishness" and "machineness". A thimac may act in the two roles simultaneously.

A thing (ignoring its mechanical nature) flows in an abstract five-dimensional structure that forms an abstract machine called a TM, as shown in Fig. 1, in which the elementary processes are called the *stages* of a TM. In the TM model, we claim that five generic processes of things exist: things can be created, processed, released, transferred, and received. These five processes form the foundation for modeling thimacs. Among the five stages, *flow* (solid arrow in Fig. 1) signifies conceptual movement from one machine to another or among the stages of a machine. The TM stages can be described as follows.

*Arrival*: A thing reaches a new machine.

*Acceptance*: A thing is permitted to enter the machine. If arriving things are always accepted, then arrival and acceptance can be combined into the *receive* stage. For simplification, the examples in this paper assume a receive stage.

*Processing* (change): A thing undergoes some kind of transformation that changes it, without creating a new thing.

*Release*: A thing is marked as ready to be transferred outside of the machine.

*Transference*: A thing is transported somewhere outside of the machine.

*Creation*: A new thing is born (created) in a machine. "Create" resembles "there is", as in, according to Whitehead [21], "*there is* Cleopatra's Needle" (italics added), and in the context of events, it resembles "*be* again" (italics added). For simplification, we may omit or keep *create*, as illustrated in Fig. 2.

In addition, the TM model includes memory and triggering (represented as dashed arrows) relations among the processes' stages (machines).

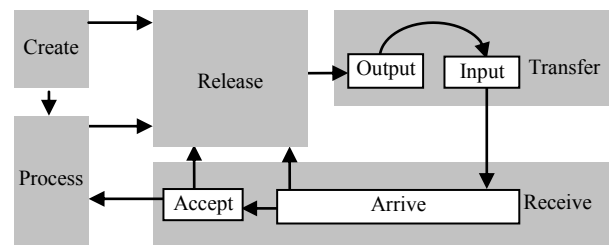

**Figure 1. A thinging machine.**

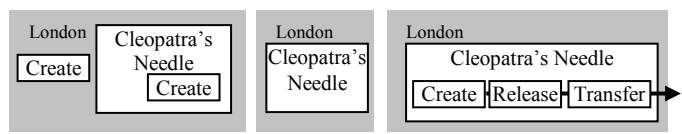

**Figure 2. Simplification by omitting *create* (middle). In some situations, *create* is necessary to indicate the origin of flow (right).**

A machine creates, in the sense that it "finds/originates" a thing; that is, it brings a thing into the system and then becomes aware of it. Creation can be used to designate "bringing into existence" in the system because what exists is what is found.

### IV. ILLUSTRATIONS OF USING THE THINGING MACHINE IN REPRESENTATION

#### A. Actions and Events

In his study of actions, Davidson [29] observed that events in a person's life reveal agency and distinguish his or her actions. "A person's actions define some events, whereas other events simply befall upon a person". For Davidson [29], "events are perceptible"; that is, in a TM, events are things that can be transferred, received, and processed. In addition, for Davidson [29], "events can be located in space and time". In a TM, events can be time and (conceptual) space thimacs in addition to other types of thimacs (e.g., intensity).

Davidson [29] gave the following example of actions:

This morning I was awakened by the sound of someone practicing the violin. I dozed a bit, then got up, washed, shaved, dressed, and went downstairs, turning off a light in the hall as I passed. I poured myself some coffee, stumbled on the edge of the dining room rug, and spilled my coffee fumbling for the *New York Times*. [29]

Davidson [29] stated that a fairly definite subclass of *events* are *actions*, and a person named as subject or object in sentences may or may not be the agent of the recorded *event*.

#### A. Static Thinging Machine Description

We convert Davidson's example into a TM to illustrate modeling in thinging machine model and to present the notions of thing, machine, and event. Minor modifications are made to the example, such as intentionally tripping over a rug, even though these additions can easily be expressed in TM. Fig. 3 shows the TM's static representation, for which we assume that the description was generated *this morning* and that it is still this morning (circle 0).

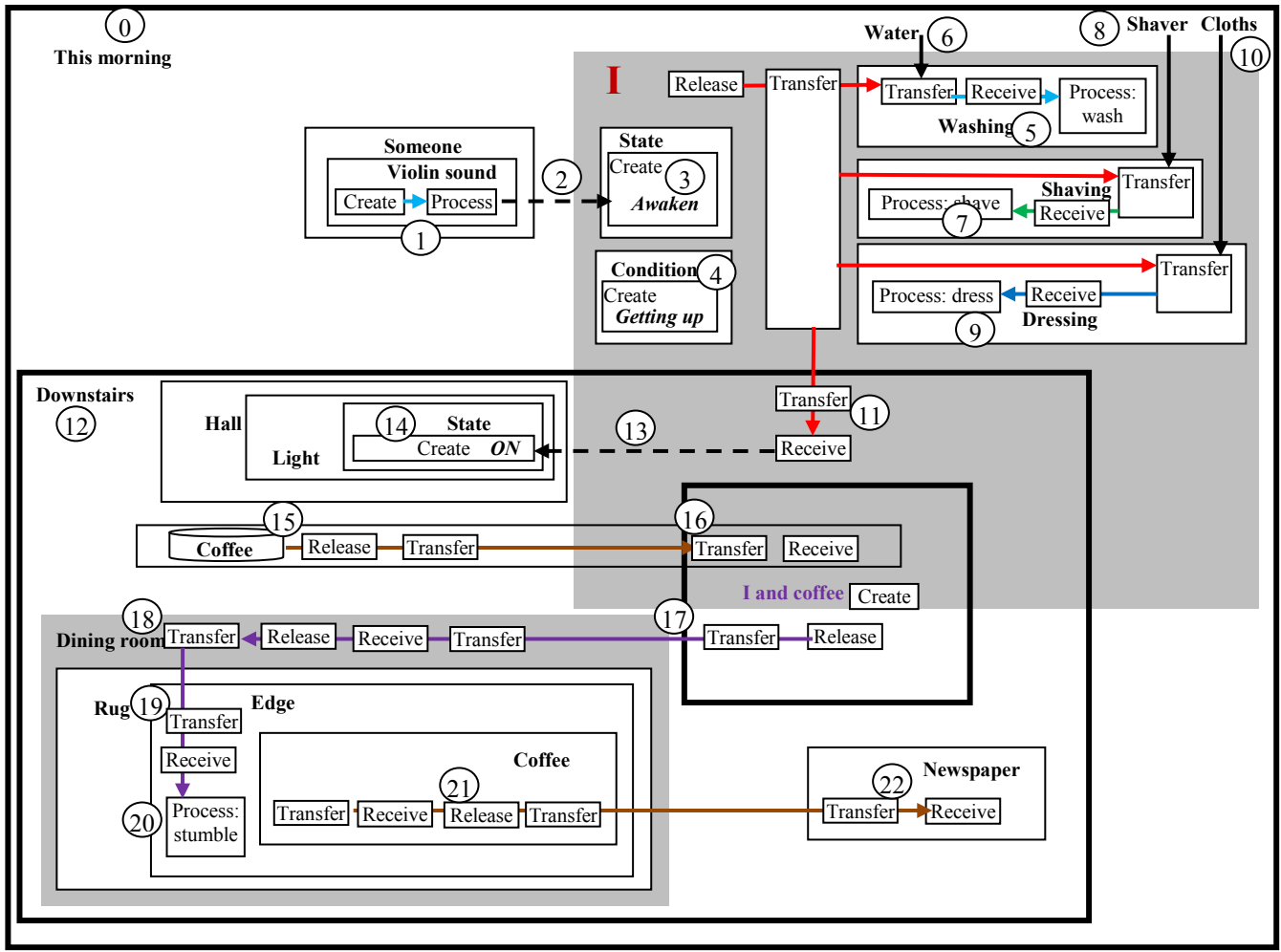

**Figure 3. TM model of Davidson's example.**

In the figure, someone processes (practices) his or her violin (1), which triggers (2) me to be in the awakened state (3), and so I get up (4). Then, I wash up (5). Here, water (6) is included in the diagram because it seems to be necessary for the washing. From the TM perspective, washing is a (conceptual) machine that involves me and water. Note that I change from having the role of a machine that is triggered to wake up to having the role of a thing that flows to a washing machine. In the same way, I enter the shaving process (7), which also involves a shaving machine (8) and a clothing machine (9) that includes clothes (10). Then, I move (11) downstairs (12), trigger (13) turning the light on (14), and pour myself coffee (15 and 16). With my coffee, I go (17) to the living room (18), and, on the edge of the rug (19), I stumble (20), spilling my coffee (21) on the newspaper (22).

### B. The System's Behavior

To specify the behavior embedded in this example, we need to use events. A TM event is a thimac that includes a time thimac. Elementary TM events can be constructed upon the generic TM stages. However, these events are too minor, from the common sense aspect, to be considered meaningful.

For example, release and transfer are usually combined and called *send*. Accordingly, identifying "meaningful" events is a design question. Fig. 4 shows a complete model of the event *I was awakened by someone practicing violin so I got up* as an event. For simplification's sake, we will represent events by their regions.

Accordingly, Fig. 5 shows the following events.

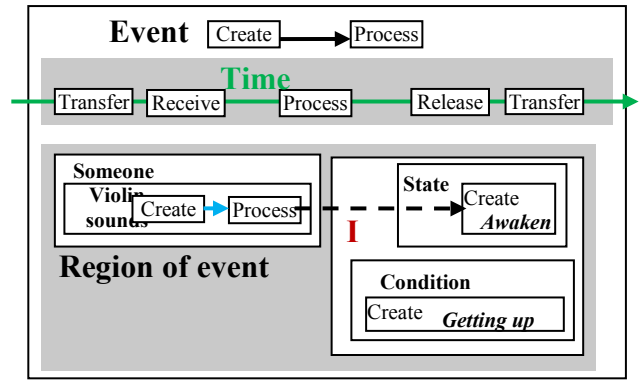

**Figure 4. The event *I was awakened by someone practicing violin so I got up*.**

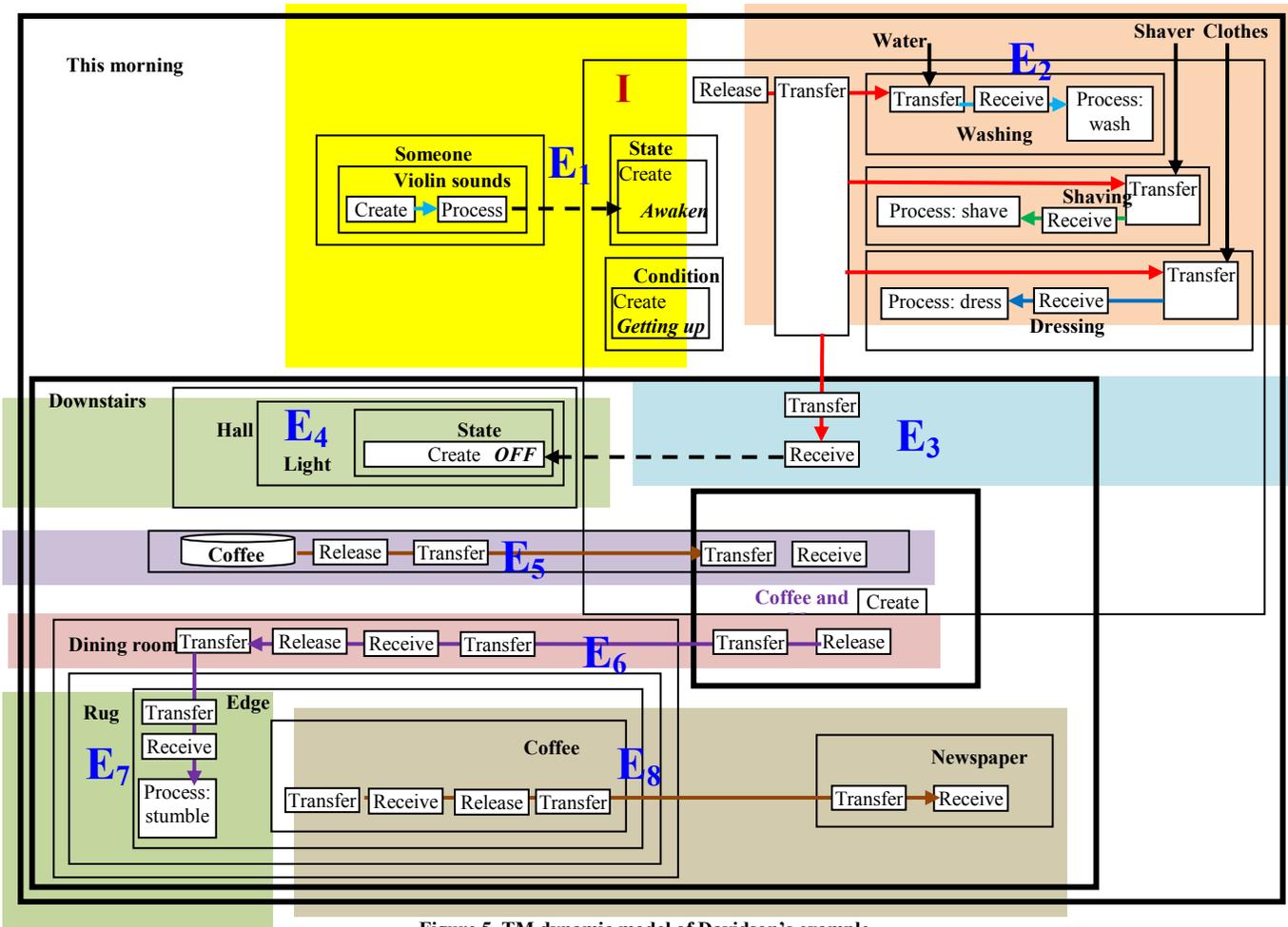

**Figure 5. TM dynamic model of Davidson's example.**

*Event 1 (E₁)*: I was awakened by someone practicing the violin, so I got up.

*Event 2 (E₂)*: I wash, shave, and dress myself.

*Event 3 (E₃)*: I go downstairs.

*Event 4 (E₄)*: I turn off a light in the hall.

*Event 5 (E₅)*: I pour myself some coffee.

*Event 6 (E₆)*: I go to the dining room.

*Event 7 (E₇)*: I stumble on the edge of the dining room rug.

*Event 8 (E₈)*: I spill my coffee while fumbling for the *New York Times*.

Fig. 6 shows the behavioristic description in terms of the chronology of these events. Of course, some of the events may occur in parallel (e.g., going downstairs and turning off the light).

From these TM descriptions, Davidson's [29] assertion that "actions are events" seems correct, in the sense that an action refers to the TM (conceptual) *regions* of events. Elementary actions in a TM are its stages.

Accordingly, the TM model can be used in this type of study. The following example from Davidson [29] illustrates this point. Davidson [29] is interested in the notion of recurrence when he states, "Here are some things that recur or happen more than once. . . . Last night I dropped a saucer of mud, and tonight I did it again (exactly the same thing happened). The 'it' of 'I did it again' looks for a reference, a thing that can recur". Fig. 7 shows how this is modeled in a TM. In Fig. 7, E represents the same event region (a thing that recurs): *I dropped a saucer of mud* recurs without direct reference to time (last night and tonight) except but preserving the sequence relationship. The figure represents the repeatability corresponding to "last night I dropped a saucer of mud, and tonight I did it again" of "tonight, I dropped a saucer of mud, and last night, I did so too".

Events can occur in sequence or parallel. It is also possible to have events of events. Consider the following sentence given by Krifka [36]: "Four thousand ships passed through the lock last year. A ship passing through the lock (which takes time)" is the event $E_{passing}$ that happened at a certain time last year. Accordingly, we can say that the event $E_{last\ year}$: "is $E_{passing}$ occurred 4,000 times" (see Fig. 8).

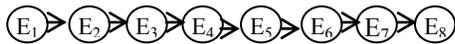

**Figure 6. Chronology of events in Davidson's example.**

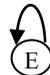

**Figure 7. Recurrence of the event *I dropped a saucer of mud*.**

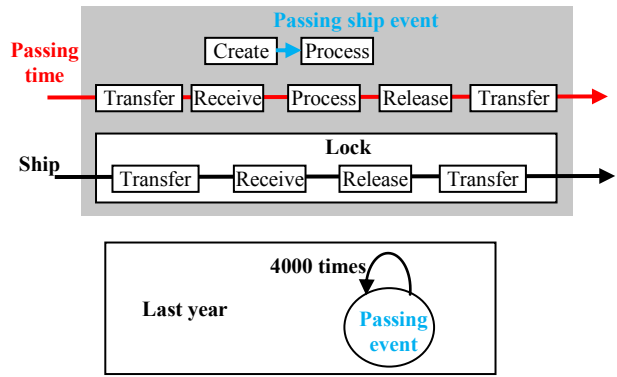

**Figure 8. The event *A ship passed through the lock* (top diagram) and the event *Four thousand ships passed through the lock last year* (bottom diagram).**

*B. Facts and Events*

According to the *Stanford Encyclopedia of Philosophy* [20], events are contrasted with objects, insofar as both are conceived of as individual and concrete entities organized into part–whole hierarchies. Some conceptions view events and objects as closely tied so as to be structurally complementary, in that any characterization of the event concept yields a characterization of the object concept through a simple replacement of temporal with spatial predicates, and vice versa:

> From this point of view, events are to be distinguished from facts, which are characterized by features of abstractness and a-temporality: the event of Caesar's death took place in Rome in 44 B.C., but that Caesar died is a fact here as in Rome, today as in 44 B.C. One could indeed speculate that for every event there is a companion fact (the fact that the event took place), but the two would still be categorically distinct. [20]

Fig. 9 shows the TM representation of the event of Caesar's death, which took place in Rome in 44 B.C. The figure expresses that *there is* an event (circle 1) that happened (create) and took its course (process) in 44 B.C. (2), and its content (the conceptual region; 3) is in the room (4) where Caesar walked to his death (5-6).

Fig. 10 shows the fact that *Caesar died here as in Rome, today as in 44 B.C*. In the figure, the fact (circle 1) appears (created) and can flow (to another time period, another person, etc.). It is a fact because its truth value is true (2). It is a fact about an event (30).

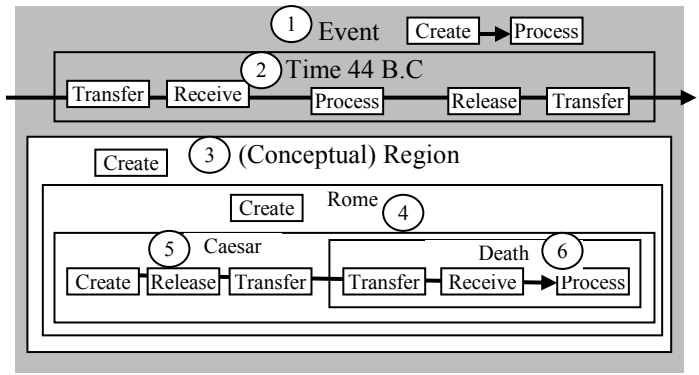

Figure 9. The event *Caesar's death took place in Rome in 44 B.C.*

## V. CASE STUDY: DYNAMIC MODELING

In object-oriented methodology, a scenario is a sequence of events that occurs during one particular execution of a system [1]. The dynamic model is graphically represented by state diagrams, in which a state corresponds to the interval between two events received by an object and describes the "value" of the object for that time period [1].

### A. Automated Teller Machine Model

Nath [1] provided the following scenario, in which a user withdraws money from an automated teller machine (ATM). Each event transmits information from one object to another.

*1)* The ATM asks the user to insert a card. The user inserts a cash card.
*2)* The ATM accepts the card and reads its serial number.
*3)* The ATM requests a personal identification number (PIN). The user enters 1234.
*4)* The ATM verifies the serial number and PIN with the consortium. The consortium checks it with bank ABC and notifies the ATM of acceptance.
*5)* The ATM requests an amount. The user enters the amount.
*6)* The ATM processes the request and dispenses the required amount of money. [1]

Fig. 11 shows the TM model for this scenario of events.

- The ATM sends the user a request to insert a card (circle 1).
- The user inserts a cash card (2) that flows into the ATM (3).
- The ATM processes (4) the card and extracts its serial number (5). Note that this extraction is modeled in terms of triggering the transfer and receipt of embedded data; that is, the serial number is received when the card is received.
- The ATM creates a request for the PIN (6) that flows to the user (7), who enters it (8). The PIN flows to the ATM (9).

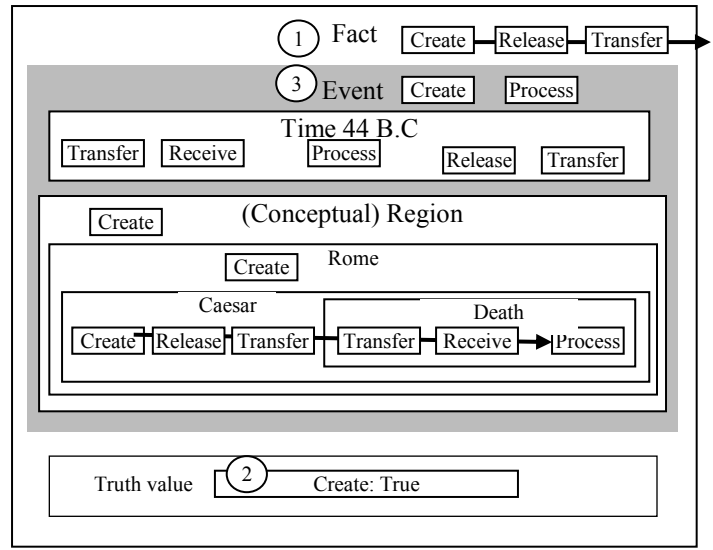

Figure 10. The fact *Caesar died here as in Rome, today as in 44 B.C.*

- The ATM constructs a request for verification (10) that contains the serial number (11) and PIN (12) and sends the request to the consortium (13).
- The consortium transmits the request to the bank (14).
- The bank processes the request (15) to extract the serial number (16) and PIN (17). A verification function (18) validates the sequence number and PIN by comparing them (19) with a database of tuples (20) of valid data (of serial number and PIN).
- Assuming that the received data are valid, an acceptance response is constructed (21) that flows to the consortium (22), which sends it to the ATM (23).
- In the ATM, the received acceptance triggers (24) a request (25) for the user to enter the amount he or she wishes to withdraw (26).
- The user enters (27) the amount, which flows (28) to the ATM, where it, along with the serial number (29 and 30), is sent to the bank via the consortium (31).
- The bank checks whether the balance is sufficient (32) and sends a message that the balance is sufficient for a withdrawal (33).
- In the ATM, the OK message is processed (34) to trigger (i) releasing the money to the user (35), (ii) sending a receipt to the user (36), and (iii) releasing the card (37).

### B. On Complexity

Model complexity is an important factor in software engineering. According to Holt [37], modeling can reduce the impacts of complexity, lack of understanding, and poor communication in project development. A model's quality depends on the quality of specifications that are suitable for development and refinement of the design at all stages.

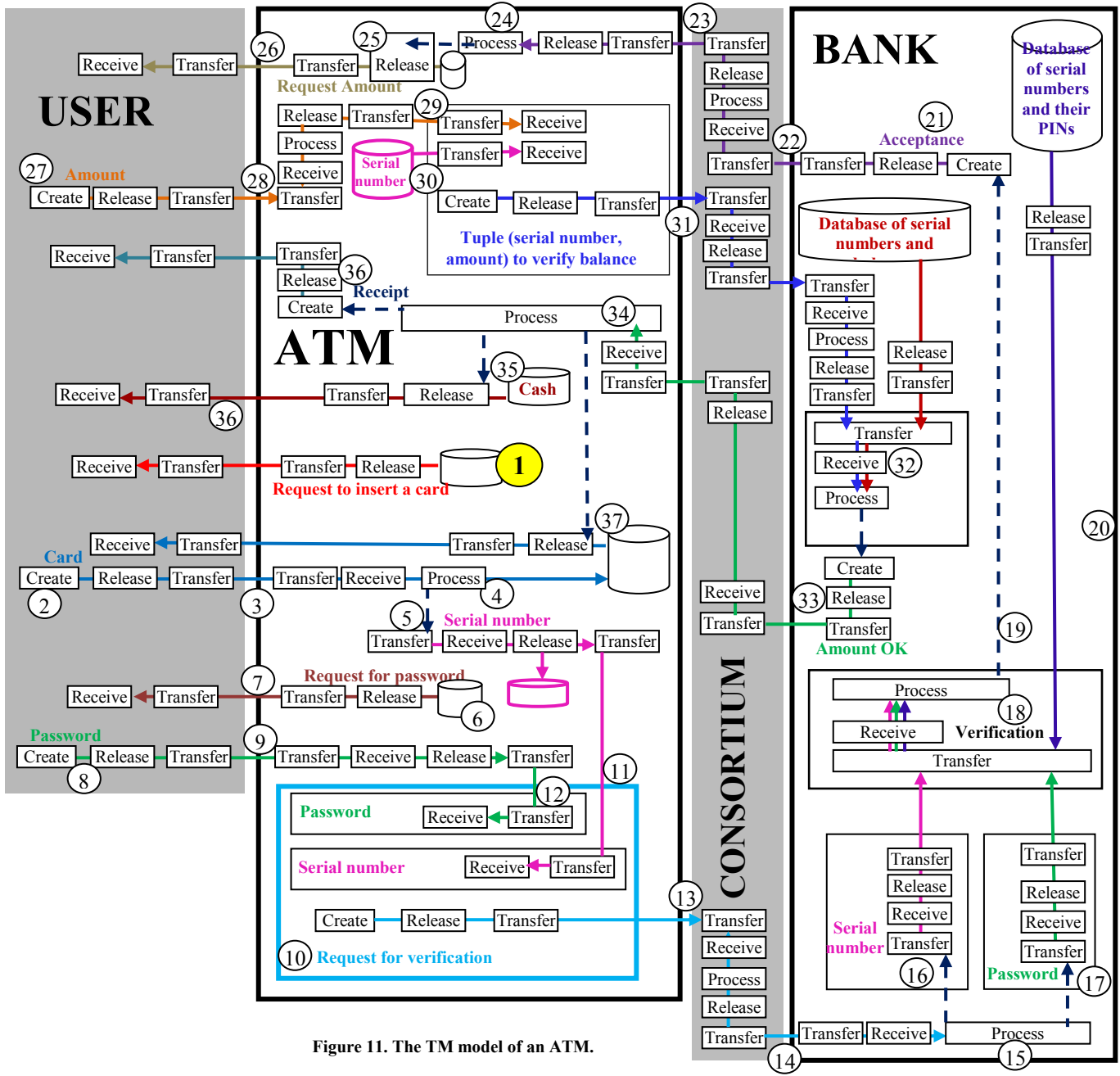

**Figure 11. The TM model of an ATM.**

One may claim that such a TM diagram is a complex representation, in comparison with the descriptions utilized in object-oriented modeling. Researchers typically assert that the object-oriented model's advantages include simulating a designer's way of thinking [38] and that "different kinds of object-oriented languages share the common feature of reducing *complexity* in the representation of technical systems and design processes" [39]. According to Duckham [40], "This proliferation has not always been complemented by a growth in [object-oriented] theory. The surfeit of object-oriented analysis, design and programming techniques which exist are, therefore, necessarily highly subjective."

As Joque [41] observed:

Despite the obvious allusion to object-oriented programming in the naming of object-oriented ontology, there are few descriptions of the relationship between object-oriented programming and said ontology. This is especially unfortunate as the history and philosophy that surround object-oriented programming offer a nuanced understanding of objects, their ability to hide part of themselves from the world, their relations, and their representation in languages that in many ways challenge the claims offered by object-oriented ontology.

According to Bishop [42], "There are multiple senses of complexity. For instance, something can be considered complex because it is complicated or intricate. However, systems that have a complicated set of interacting parts may actually exhibit relatively simple behavior."

One interesting aspect of the TM model is its systematic application of the five generic stages. The repeatability of this application creates more detailed specifications; however, a TM diagram may be simplified to any required level of granularity, based on the original TM description. For example, Fig. 12 was produced from the static TM representation of an ATM. The release, transfer, and receive stages were removed, based on the assumption that the arrow direction is enough to indicate the flow of things.

### C. Event-Trace Diagram

According to Nath [1], the ATM scenario's limitations are that it is unclear how many objects are involved, which objects generate an event, and which objects receive an event. Nath [1] developed an event-trace diagram (sequence diagram) that can show both the sequence of events and the objects exchanging events.

In the sequence diagram, each object is a vertical line, and each event is a horizontal arrow from the sender object to the receiver object. Time increases from top to bottom (see Fig. 13).

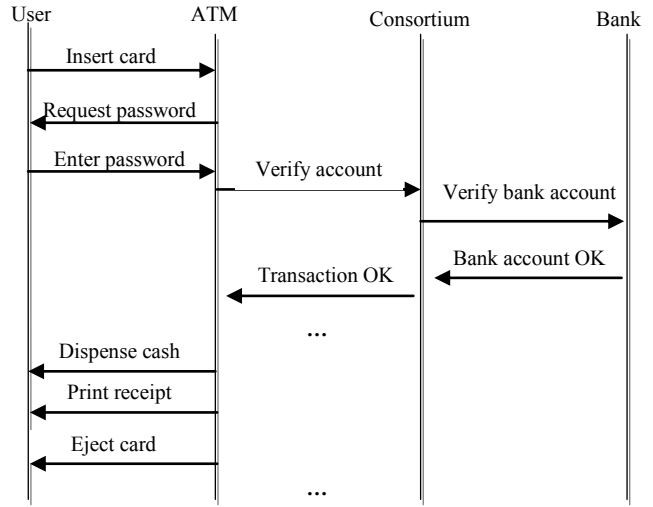

Figure 13. Sequence diagram (redrawn, partially from [1]).

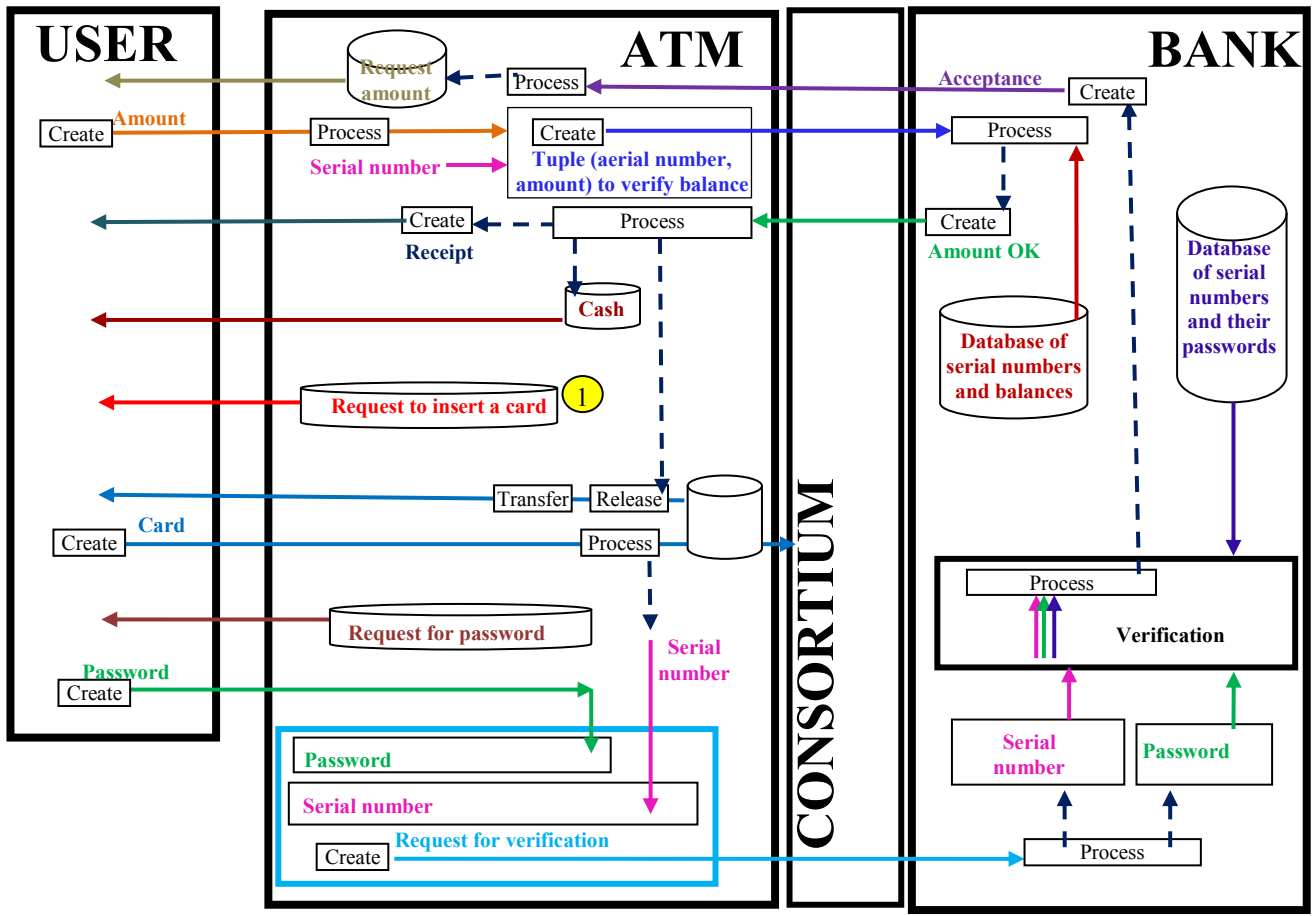

**Figure 12. Simplification of the TM model for an ATM by eliminating release/transfer and transfer/receive, under the assumption that the arrows' directions are sufficient to indicate the flow of things.**

Fig. 14 shows a diagram of the TM events for an interaction with the ATM transaction. We select the following events in building the system's behavior model.

*Event 1 ($E_1$)*: The ATM asks the user to insert his/her card.

*Event 2 ($E_2$)*: The user inserts a cash card that flows into the ATM, to be processed to extract the card's serial number.

*Event 3 ($E_3$)*: The user is asked to input his or her PIN.

*Event 4 ($E_4$)*: The user inputs the PIN.

*Event 5 ($E_5$)*: A request for validation that includes the serial number and password is created.

*Event 6 ($E_6$)*: The validation request flows to the bank via the consortium.

*Event 7 ($E_7$)*: The bank extracts the serial number and PIN and checks them with its database.

*Event 8 ($E_8$)*: An acceptance response is created and sent to the ATM via the consortium.

*Event 9 ($E_9$)*: The ATM asks the user to input the transaction amount, assuming it is a withdrawal.

*Event 10 ($E_{10}$)*: The amount and the serial number are sent to the bank via the consortium.

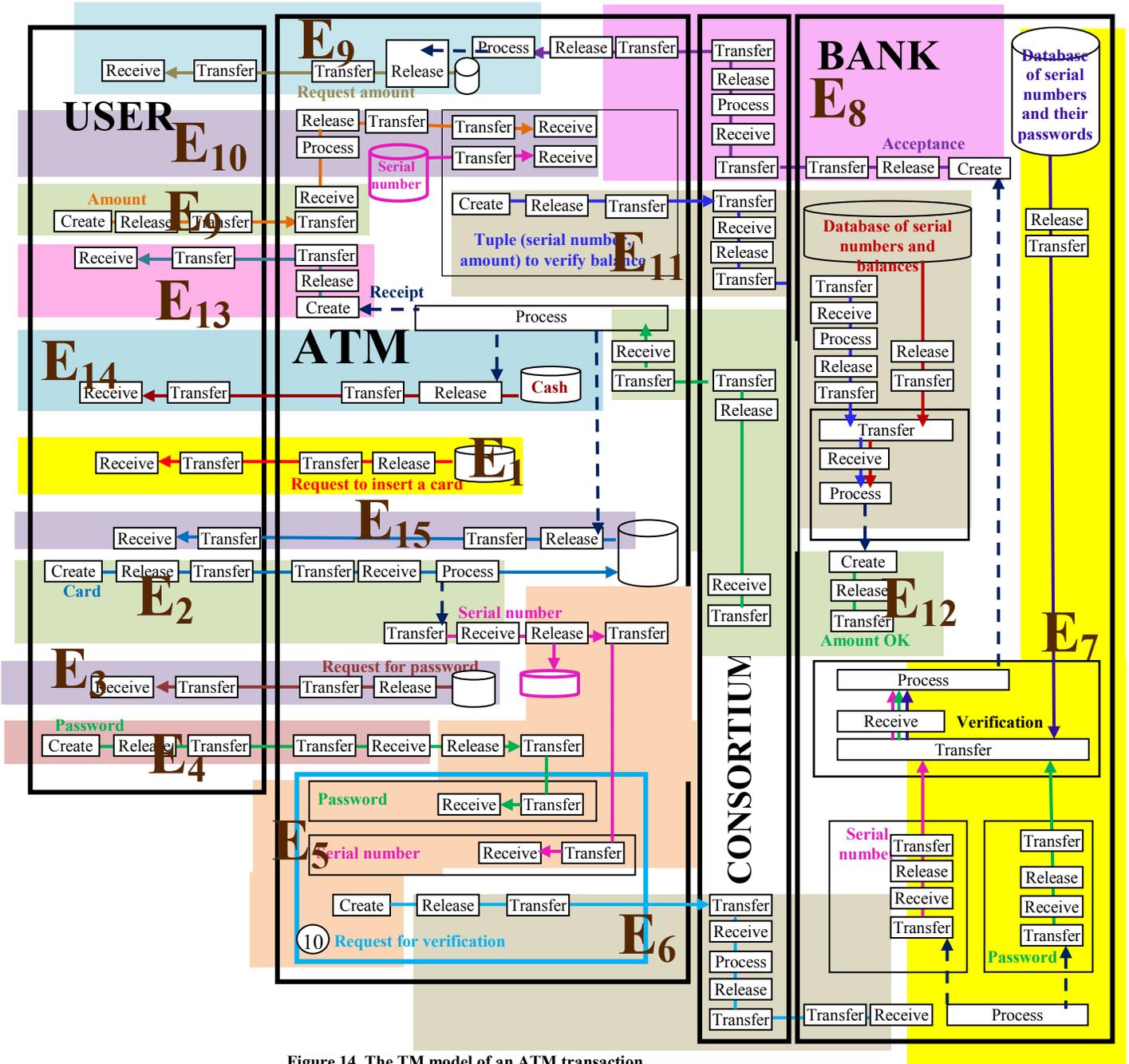

**Figure 14. The TM model of an ATM transaction.**

*Event 11 ($E_{11}$)*: The bank checks the requested amount against the user's current balance.

*Event 12 ($E_{12}$)*: The bank sends a message that the balance is sufficient.

*Event 13 ($E_{13}$)*: A receipt is provided to the user.

*Event 14 ($E_{14}$)*: Cash is dispensed to the user.

*Event 15 ($E_{15}$)*: The card is returned to the user.

Fig. 15 shows the ATM system's chronology of events. It is reasonable to observe the large notational jump in the object-oriented methodology between the static description of the system (e.g., scenario or class diagram) and the behavioral description such as the sequence diagram discussed above. It is puzzling how Nath's object-oriented requirements analysis started with events (scenario) then ended with events of the sequence diagram to (according to him) connect events to objects.

## VI. CONCLUSION

In this paper, we contrasted the object-oriented methodology with the TM model, emphasizing the notion of events in both modeling approaches. In modeling, a variety of thinking methods apply to designing the intended system [43]. Understanding events has important implications for systems modeling, and TM seems to present a more systematic approach to the definition of an event than other alternatives. In a TM, events are treated uniformly as thimacs (the basic ontological entities) that enter the model when representing behavior. The entire modeling process proceeds using the same notations throughout the stages, from the static to the dynamic specifications of the system.

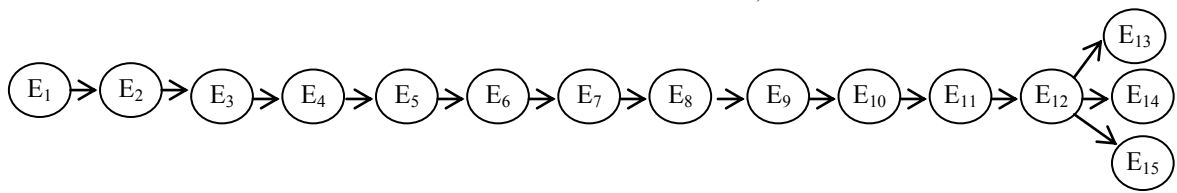

**Figure 15. The chronology of events in the ATM example.**